# A Lightweight Energy Management Method for Hybrid PV/Battery/Load Systems

Mohsen Banaei
Department of Applied Mathematics and Computer Science
Technical University of Denmark, Copenhagen, Denmark
e–mail: moban@dtu.dk

Razgar Ebrahimy, Henrik Madsen
Department of Applied Mathematics and Computer Science
Technical University of Denmark, Copenhagen, Denmark
e–mail: raze@dtu.dk, hmad@dtu.dk

## ABSTRACT

In this paper, a computationally lightweight algorithm is introduced for hybrid PV/Battery/Load systems that is price responsive, responds fast, does not require powerful hardware, and considers the operational limitations of the system. The method is applied to two buildings equipped with PV and battery. Simulation results show that the method can give results that are up to 3.9% more expensive than the Model predictive control (MPC) approach while the runtime of the program is up to 1000 times less than the MPC. Also, while the runtime of the proposed method is in the range of the self-consumption maximization (SCM) approach as the fastest method, its electricity cost is about 3.2% cheaper than the SCM method. Simulation results also show that in case of providing grid services by the battery the difference between electricity cost of the proposed approach and MPC can reduce which makes the method good for such applications.

## KEYWORDS

Energy management, Demand response, Renewable energy, Model predicative control, Optimization, Battery scheduling.

## INTRODUCTION

Global concerns about climate change and the need to reduce greenhouse gas emissions have led to rapid growth in renewable energy. In addition to their environmental benefits, renewable energy resources can offer energy security, independence, sustainability, and decentralization. Photovoltaic (PV) is one of the fast-growing renewable energy resources due to its scalability and modularity, wide geographic accessibility, low operating and maintenance costs, rapid deployment, shorter construction time, and the possibility of distributed energy production. At the moment global installed capacity of PV is around 1100 GW [1]. Despite the different advantages of PVs, their intermittent nature and dependency on solar radiation may limit the utilization of this energy resource. While peak electricity demand in buildings happens in the morning and evening, the PV panels are in their zero or low output power modes and produce energy mostly during the mid-peak hours. This mismatch can lead to power shortage or the overloading of distribution power networks and worsen the power quality [2]. Moreover, the buying and selling price of energy for households are very different and the injected power to the grid is usually settled with low tariffs, which is not cost-efficient.

To overcome this issue, PVs are recommended to be operated parallel to batteries. In this situation, optimal charging and discharging scheduling of the battery plays an important role in the cost-effectiveness of this setup. There are many research works in the literature to address this issue from different perspectives. Some studies are focused on maximizing self-consumption of the building such as the proposed methods in which the battery is charged when there is an excess of PV production and discharged when PV production is less than the building's power consumption [3]. Another group of studies applies time-of-use strategies in which one or more charging and discharging periods are defined based on the electricity prices and the battery is charged and discharged in these periods. There are also mixed self-consumption maximization with ToU arbitrage methods which combine the two strategies [3]. The abovementioned methods can be categorized as rule-based methods.

Other groups of studies propose optimal solutions for energy management of hybrid PV/Battery/Load systems. The most popular methods used to solve optimization problems are Mixed Integer Linear Programming (MILP) and Dynamic Programming (DP) [4][5]. Moreover, in some cases, MILP is applied to the Model Predictive Control (MPC) approach to include the latest updates of forecasts in energy management [6]. Other methods such as Artificial Neural Network (ANN) have also received increasing attention during the last years [7]. The rule-based methods are fast in response and do not require powerful hardware to run but their results are not optimal. On the other hand optimization methods lead to optimal results but they have the highest computational time and require powerful hardware. The ANN-based methods require a training period and their results also are not optimal.

In this paper, a lightweight energy management method is proposed for hybrid PV/ Battery/Load systems. The proposed method is supposed to fill the gap between rule- based and optimal methods, i.e., being fast in response with close to optimal results. The main application of such an algorithm can be for load aggregation and coordination mechanisms in which fast decision-making and transactions between aggregators and end-users are needed for providing grid services.

## REVIEWING THE GENERAL RULE-BASED AND MPC METHODS

The effectiveness of the proposed method is evaluated by comparing the results with rule-based and MPC methods. The block diagram of the SCM method is presented in Figure 1 [8].

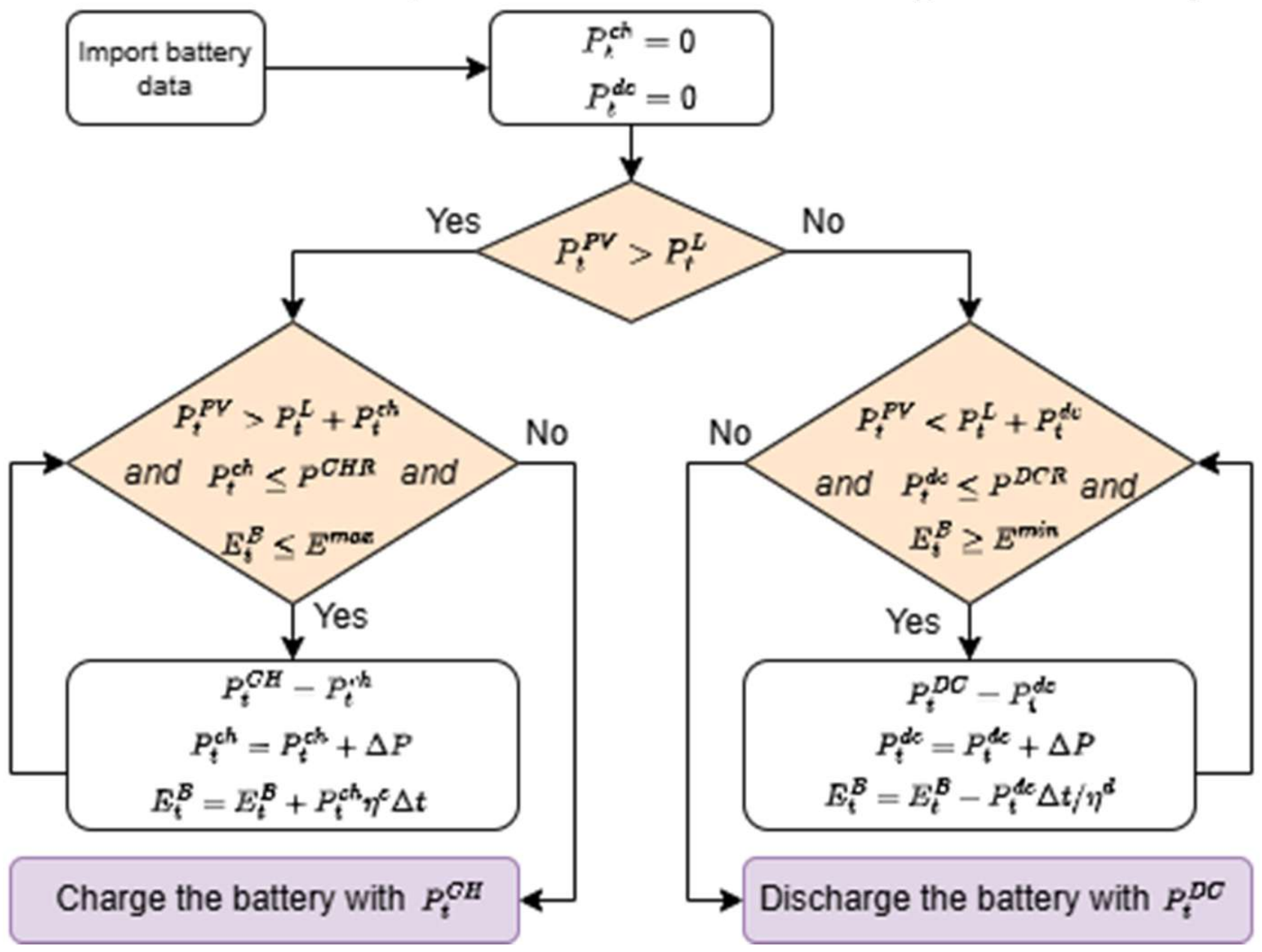

Figure 1. Block diagram of the SCM method

As shown in Figure 1 when the output power of the PV ($P_t^{PV}$) is more than the building load ($P_t^L$) the battery is charged with $P_t^{CH}$such that the maximum charging rate ($P^{CHR}$) and energy storage capacity ($E^{max} = E^N SOC^{max}$) limitations of the battery is respected. Similarly, when the output power of the PV is less than the building energy consumption, the battery is discharged with $P_t^{CH}$ considering the discharge rate ($P^{DCR}$) and minimum energy storage ($E^{min} = E^N SOC^{min}$) limit of the battery. $\eta^c$ and $\eta^d$ represent the charging and discharging efficiencies of the battery, respectively.
According to MPC approach, at each time interval, an optimization problem is solved to consider the last updates in the predictions. For PV/Battery/load systems this optimization problem is formulated as a MILP.
The MILP approach for obtaining the optimal control strategy in arbitrary time interval $t_0$ can be formulated as below [9]:

$$\min \sum_{t=t_0+1}^{t_0+T} (\rho_t^b B_t - \rho_t^s S_t)\Delta t \tag{1}$$

$$s.t.$$

$$B_t + S_t + P_t^{PV} + P_t^{DC} - P_t^{CH} = P_t^L \qquad \forall t = t_0 + 1, \dots t_0 + T \tag{2}$$

$$P_t^{DC} - P^{DCR} X_t \leq 0 \qquad \forall t = t_0 + 1, \dots t_0 + T \tag{3}$$

$$P_t^{CH} - P^{CHR} Y_t \leq 0 \qquad \forall t = t_0 + 1, \dots t_0 + T \tag{4}$$

$$X_t + Y_t \leq 0 \qquad \forall t = t_0 + 1, \dots t_0 + T \tag{5}$$

$$E_t^B = E_{t-1}^B - \Delta t\left(\frac{P_t^{DC}}{\eta^d} - \eta^c P_t^{DC}\right) \qquad \forall t = t_0 + 1, \dots t_0 + T \tag{6}$$

$$E^{min} \leq E_t^B \leq E^{max} \qquad \forall t = t_0 + 1, \dots t_0 + T \tag{7}$$

where $\rho_t^b$ and $\rho_t^s$ are energy buying and selling prices, respectively. The objective function (1) minimizes the electricity cost. $B_t$ and $S_t$ represent the purchased power from the grid and injected power into the grid, respectively. Constraint (2) represents the power balance in the system. Constraints (3)-(5) are used to satisfy charging and discharging rates constraints. Auxiliary variables $X_t$ and $B_t$are used to prevent simultaneous charging and discharging of the battery. Constraint (6) updates the energy storage level of the battery and constraint (7) defines the upper and lower bounds of the energy storage of the battery.

At each time interval, optimization problem (1)-(7) is solved, however, only the results of the first time interval are applied to the system and the rest are ignored. This process continues until the last time interval.

## INTRODUCING THE PROPOSED METHOD

The proposed method is inspired by the Packetized Energy Management (PEM) method, but unlike PEM, it is designed in such a way that is responsive to prices and includes more operational limitations [10]. the block diagram of the proposed method is illustrated in Figure 2.

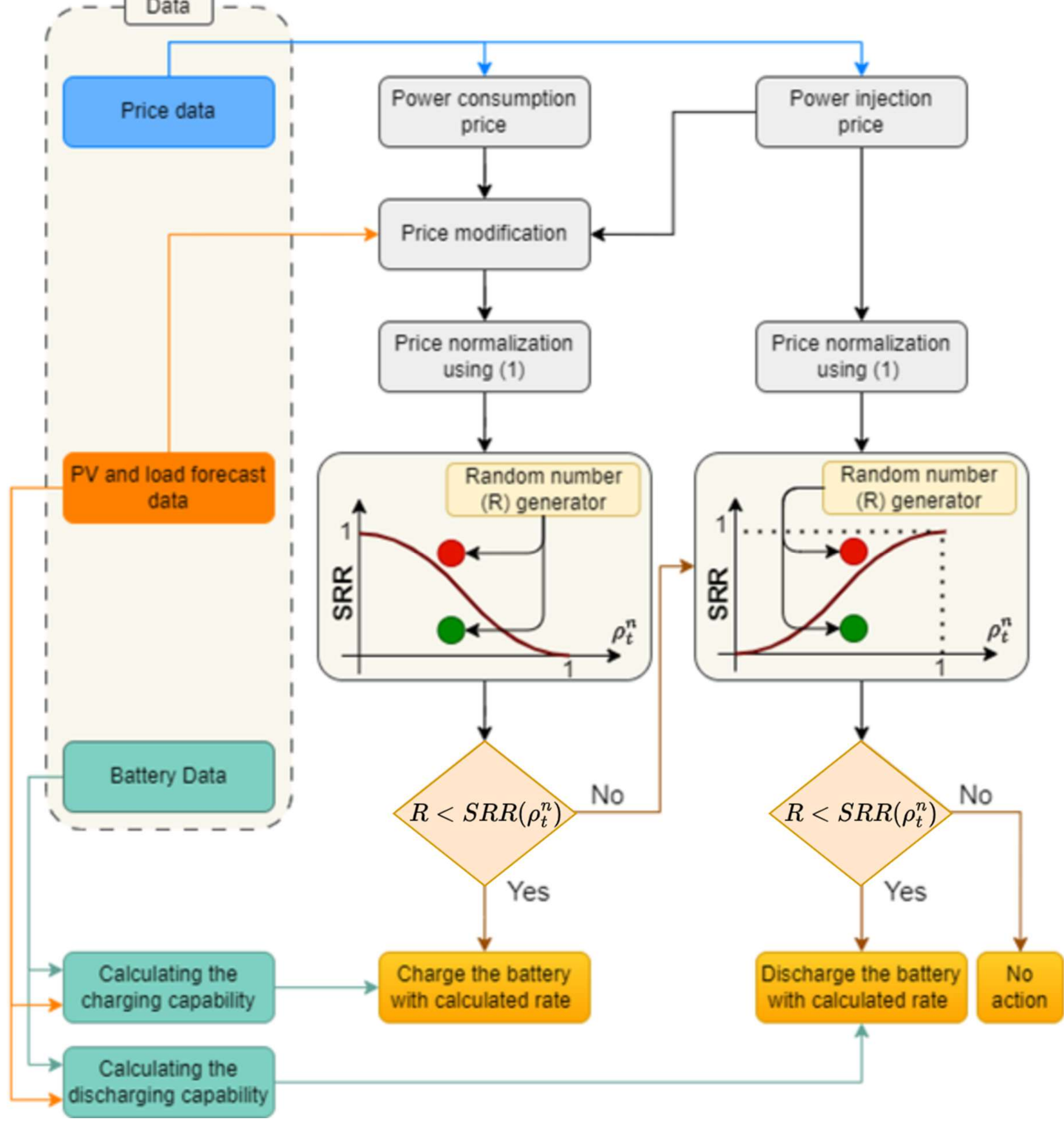

Figure 2. Block diagram of the proposed method

The method works with normalized prices. So, in the first step, the price $\rho_t$ should be normalized using below equation:

$$\rho_t^n = \frac{\rho_t - \rho^{min}}{\rho^{max} - \rho^{min}} \tag{8}$$

where $\rho^{min}$ and $\rho^{max}$ are minimum and maximum prices in the time series and $\rho_t^n$ is the normalized price.

As we know, the price of injecting power into the grid is different from the price of purchasing power from the grid. In Denmark, the price of purchasing power ($\rho_t^b$) is calculated as the day-ahead market price plus taxes and tariffs. The price of injecting power into the grid $\rho_t^s$) is equal to the day-ahead market price. So, price normalization should be done for both prices and hence, we have $\rho_t^{bn}$ and $\rho_t^{sn}$ as normalized purchasing and selling prices calculated by (8).

Battery charging occurs in two cases, 1) the price is low, and 2) PVs' output power is greater than the power consumption of the building. On the other hand, price is the main variable that affects on decision on charging or discharging the battery or taking no action.

In the first case, the normalized purchasing price can be easily used to make the decision on charging. In the second case, it is suggested to modify the price in such a way that high priority is assigned for charging when the output power of PV is greater than the energy consumption. So, the time series is checked, and for time intervals that $P_t^{PV} > P_t^L$ we assign the minimum price in the time series $\rho^{min}$ as the price. Now we normalize the modified prices and use them as the input for the core of the algorithm.

Battery discharging occurs when the price is high. So, in this case, normalized selling prices can be directly used as input for the core of the algorithm as shown in Figure 2.

The principle of the method for both charging and discharging decisions is the same. First, a stochastic request rate (SRR) function is defined. This function shows how the probability of a request for charging or discharging changes by changing the price. In the charging case, by increasing the price, the probability of sending a charging request decreases. So, the SRR function for charging mode should be a decreasing function. Also, when the normalized price is zero the value of the SRR function should be equal to 1, i.e., 100% probability of charging for minimum price in the time series, and when the normalized price is equal to 1, the value of the SRR function should be equal to zero, i.e., 0% probability for charging for maximum price. Taking into account above explanations, the SRR function for charging is formulated as below:

$$SRR^{CH} = 1 - e^{-k^{ch}(1-\rho_t^{bn})/(\rho_t^{bn}+\varepsilon)} \tag{9}$$

where $k^{ch}$ is a design parameter and $\varepsilon$ is a very small positive number.

The SRR function for discharging is an increasing function. When the normalized price is low the probability of sending a discharge request is zero, but for high prices, this probability increases. The SRR function for discharging is suggested to be formulated as below:

$$SRR^{DC} = 1 - e^{-k^{dc}\rho_t^{sn}/(1-\rho_t^{sn}+\varepsilon)} \tag{10}$$

where $k^{dc}$ is a design parameter.

To decide on charging or discharging, the algorithm generates a random number ($R$) with uniform distribution between zero and 1. If $R$ is less than the value of the SRR function for charging at the current normalized modified price ($SRR^{CH}(\rho_t^{bn})$), the charging request will be sent, otherwise, the possibility of discharge request is checked. Similarly, a random number $R$ is generated and compared with the SRR function for discharging at the current normalized price($SRR^{DC}(\rho_t^{sn})$). if $R < SRR^{DC}(\rho_t^{sn})$ a discharge request will be sent, otherwise, no action will be taken.

After determining the charging or discharging decision, the amount of the power that should be charged or discharged should also be calculated. The charging rate is limited to three different factors: 1) the nominal charging rate of the battery, 2) the maximum energy that can be stored in the battery, and 3) the extra output power of PV after supplying the load. So, the charging rate is calculated as below:

$$P_t^{CH} = \begin{cases} \min\{P^{CHR}, \dfrac{E^{max} - E_t^b}{\eta^c \Delta t}\} & P_t^{PV} \le P_t^L \\ \min\{P^{CHR}, \dfrac{E^{max} - E_t^b}{\eta^c \Delta t}, P_t^{PV} - P_t^L\} & P_t^{PV} > P_t^L \end{cases} \tag{11}$$

According to (12), when the output power of PV is less than the building's energy consumption, the charging rate is calculated as the minimum of the nominal charging rate of the battery and the free capacity in the battery for charging (preventing from overloading the battery). When the output power of PV is greater than building's load, the extra power generation of PV is also considered in the calculations (covering the power mismatch between PV and loads).
Similarly, the discharging rate of the battery is calculated as below:

$$P_t^{DC} = \begin{cases} \min\{P^{DCR}, \eta^d \dfrac{E_t^b - E^{min}}{\Delta t}\} & P_t^L \leq P_t^{PV} \\ \min\{P^{DCR}, \eta^d \dfrac{E_t^b - E^{min}}{\Delta t}, P_t^L - P_t^{PV}\} & P_t^L > P_t^{PV} \end{cases} \quad (12)$$

## CASE STUDY

The test system is a set of two buildings (A and B) in Sønderborg in South Jutland, Denmark. Each building is equipped with PV panels and a battery. Real energy consumption, PV power production, and battery data of these building are used to evaluate the performance of the proposed method. It is assumed that the PVs' output power and buildings' energy consumption are predicted precisely and the historical data are used as input for all methods. Similar batteries are used at each building. Characteristics of the battery are presented in Table 1.

Table 1. Characteristics of the battery

| $E^N$ | $P^{CHR}$ | $P^{DCR}$ | $\eta^c$ | $\eta^d$ | $SOC^{max}$ | $SOC^{min}$ | $SOC^0$ |
|---|---|---|---|---|---|---|---|
| 13.5 (kWh) | 7(kW) | 7 (kW) | 0.97 | 1 | 0.9 | 0.1 | 0.3 |

The Nordpool day-ahead market prices are used as dynamic prices. The fixed taxes and tariffs that are added to the day-ahead market price are assumed to be 0.2 (Euro/kWh). In order to evaluate the proposed method in different conditions, simulations are repeated for one month of each season, i.e., January, May, July, and October.
Design parameters $k^{ch}$ and $k^{dc}$ are assumed to be 0.3. This value is chosen by trying a wide range of numbers, however, in practical cases, more advanced methods should be used to find optimal values for these parameters.

### Battery scheduling results

To show how the proposed method works, its scheduling results are compared with SCM and MPC methods for a period. Simulation results for the first five days of May, 2022, for building A are presented in Figure 3. It can be seen that in SCM method the charging happens only when there is excess PV production and the battery is discharged right after the PV production gets less than the building's energy consumption. On the other hand, both MPC and the proposed method not only react to the PV production status but also to the prices, hence, the number of charging and discharging sequences is much more than the SCM method.

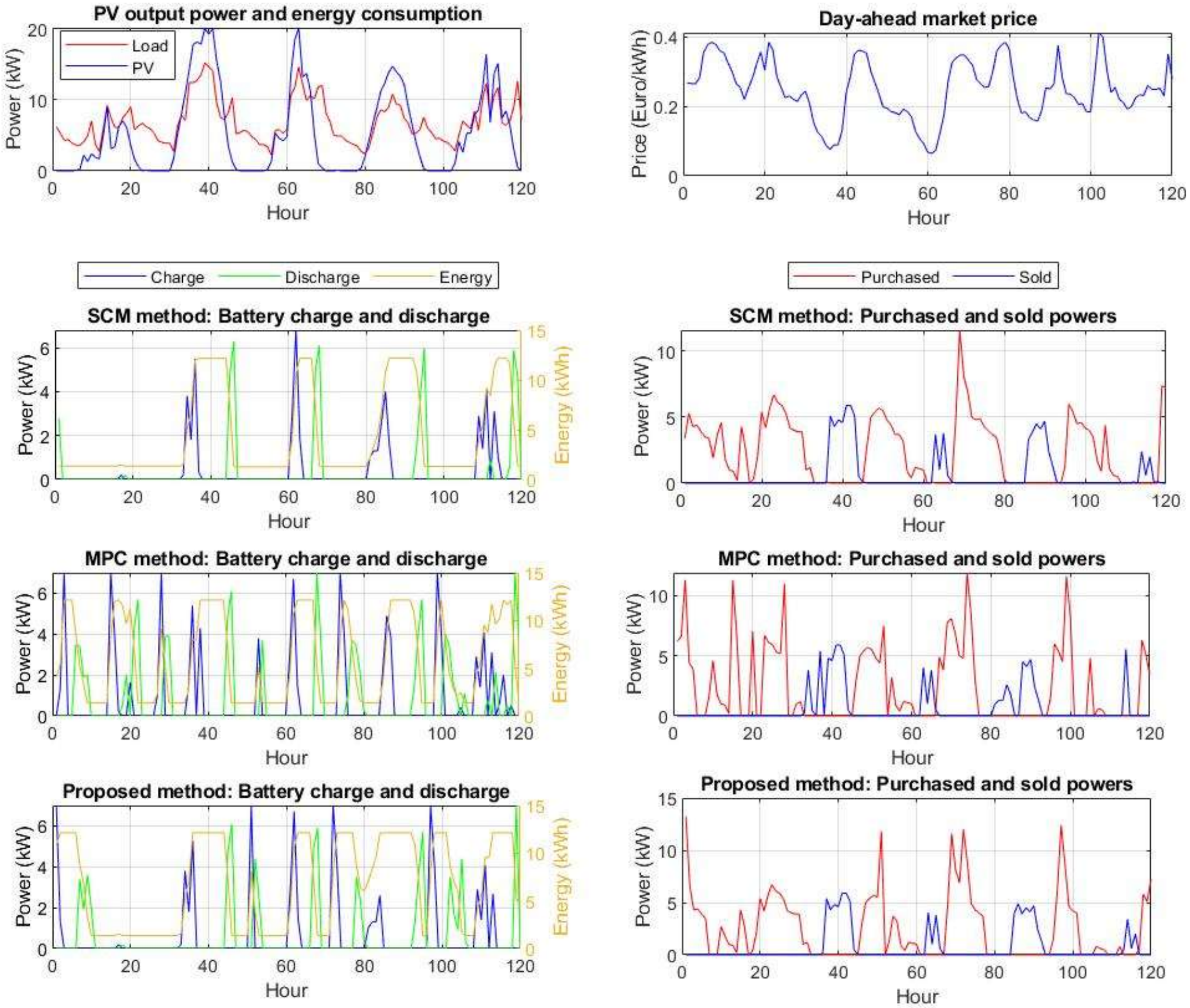

Figure 3. Battery scheduling results for SCM, MPC and the proposed method

**Analysing the cost-effectiveness and computation time of the methods**

In this section, the electricity cost and runtime of the proposed method are compared with the SCM method as a fast approach and the MPC method as an optimal method for both buildings A and B. Simulations are performed for four months January, May, July, and October to evaluate the results in different situations. The total electricity cost and runtime of the code for each approach are presented Table 2 and Table 3. Since each method includes design parameters that affect the results, the simulations are performed for different parameters of methods. In the SCM method, parameter $\Delta P$ (see Figure 1) and in the MPC method control horizon parameter $T$ affect the results. So, simulations are repeated for different values of these parameters.

Table 2. Comparing the electricity cost and runtime of the code for different methods applied to building A.

| | SCM method | | | MPC method | | | Proposed method |
|---|---|---|---|---|---|---|---|
| | $\Delta P = 0.1$ | $\Delta P = 0.5$ | $\Delta P = 1$ | $T = 8$ | $T = 16$ | $T = 24$ | |
| Cost (Euro) | 2160.22 | 2139 | 2115 | 1991.2 | 1991.1 | 1991.1 | 2070 |
| Runtime (s) | 0.95 | 0.31 | 0.26 | 577.2 | 481.7 | 312.9 | 0.31 |

Table 3. Comparing the electricity cost and runtime of the code for different methods applied to building B.

| | SCM method | | | MPC method | | | Proposed method |
|---|---|---|---|---|---|---|---|
| | $\Delta P = 0.1$ | $\Delta P = 0.5$ | $\Delta P = 1$ | $T = 8$ | $T = 16$ | $T = 24$ | |
| Cost (Euro) | 2809.2 | 2804.1 | 2804.2 | 2652.9 | 2652.3 | 2652.3 | 2720.01 |
| Runtime (s) | 0.47 | 0.15 | 0.12 | 268.7 | 527.3 | 581.3 | 0.43 |

As shown in Table 2, the runtime of the proposed method is in the range of the SCM method while it gives around 3.2% cheaper scheduling results. On the other hand, MPC method results are 3.9% less expensive than the proposed method but their runtime in the best case is about 1000 times more than the runtime of the proposed method. Similarly, for building B, the electricity cost of the proposed method is 3% cheaper than the SCM method and 2.5% more expensive than the MPC while its runtime is at least 623 times less than the MPC.

Reviewing the results in Table 2 and Table 3 shows that the electricity price structure in Denmark is such that the use of advanced and complex methods does not lead to a significant decrease in the electricity bills of end-users. This is mostly due to the fixed taxes and tariffs added to the day-ahead electricity market. Replacing the existing fixed tariffs with variable tariffs will provide a good incentive for end-users to move toward more advanced approaches.

**Impacts of applying external signals on the cost-effectiveness of the method**

Flexible loads can be in contract with aggregators for providing grid services. In this case, the aggregator may send signals to request for charging or discharging of the battery to provide grid services such as voltage regulation. Applying these external signals deviates the system operation from its optimal point. In this section, we study the impacts of these external signals on the cost-effectiveness of the proposed approach. To this end, it is assumed that random charging or discharging signals with specific probabilities are generated and sent to the batteries. The battery interrupts its routine and responds to the request. Simulation results of the cases with different probabilities of sending external signals to building B are presented in Table 4.

Table 4. Comparing the electricity cost for different probabilities of sending external signals to building B.

| | Probability of receiving external signal | | | |
|---|---|---|---|---|
| | 0% | 10% | 20% | 30% |
| MPC | 2652.3 | 2738.2 | 2844.06 | 2920.6 |
| Proposed method | 2720.01 | 2803.22 | 2903.06 | 2960.6 |
| % of difference | 2.521% | 2.373% | 2.072% | 1.369% |

As shown in Table 4, as the probability of receiving an external signal increases the difference between the electricity cost of the MPC method and the proposed method decreases. When the probability of receiving external signals reaches 30%, the difference reduces to 1.36% which is not significant.

## CONCLUSION

In this paper, a lightweight energy management method for hybrid PV/Battery/load systems is proposed. The method uses price as input and a stochastic process for deciding on charging, discharging, or taking no action. In order to evaluate the performance of the method, its results are compared with a simple rule-based method known as SCM approach and a more complex and optimal approach known as MPC.
Simulation results show that the proposed method yields scheduling results that are less than 3.9% more expensive than the MPC approach but with a runtime of up to 1000 times less than the MPC and close to the SCM method.
Another case was also studied in which an external party, for example, an aggregator, can modify the battery operation by sending charging or discharging signals for providing grid services. Simulation results show that in this case, the difference between the electricity costs of the proposed method and MPC reduces. This makes the proposed method a good alternative for such applications.
From the end-user perspective, they can install an energy management device that is price responsive and does not increase their electricity bill compared to an energy management device with MPC, significantly, but it is much cheaper. This can be a good incentive for end-users to join the demand response and flexibility services programs.

## ACKNOWLEDGEMENT

This research was supported by the European Commission through the H2020 project Ebalance-plus (Grant No. 864283).

## REFERENCES

1. Global installed pv capacity, https://www.pv-tech.org/global-installed-pv-capacity-passes-1-18tw-iea/, [Accessed: 2023-09-06] (2023).
2. Y. Yang, H. Wang, F. Blaabjerg, T. Kerekes, A hybrid power control concept for pv inverters with reduced thermal loading, *IEEE Transactions on Power Electronics 29* (12) (2014) 6271–6275
3. D. Azuatalam, K. Paridari, Y. Ma, M. F¨orstl, A. C. Chapman, G. Verbiˇc, Energy management of small-scale pv-battery systems: A systematic review considering practical implementation, computational requirements, quality of input data and battery degradation*, Renewable and Sustainable Energy Reviews 112* (2019) 555–570
4. M. Banaei, B. Rezaee, Fuzzy scheduling of a non-isolated micro-grid with renewable resources, *Renewable Energy 123* (2018) 67–78
5. Y. Li, J. Peng, H. Jia, B. Zou, B. Hao, T. Ma, X. Wang, Optimal battery schedule for grid-connected photovoltaic-battery systems of office buildings based on a dynamic programming algorithm, *Journal of Energy Storage 50* (2022) 104557
6. J. Hu, Y. Xu, K. W. Cheng, J. M. Guerrero, A model predictive control strategy of pv-battery microgrid under variable power generations and load conditions, *Applied Energy 221* (2018) 195–203
7. L. Barelli, G. Bidini, F. Bonucci, A. Ottaviano, Residential micro-grid load management through artificial neural networks, *Journal of Energy Storage 17* (2018) 287–298.
8. W. Xiangqiang, Z. Tang, D.-I. Stroe, T. Kerekes, Overview and comparative study of energy management strategies for residential pv systems with battery storage, *Batteries 8* (2022)
9. M. Banaei, F. D'Ettorre, R. Ebrahimy, E. M. V. Blomgren, H. Madsen, Mutual impacts of procuring energy flexibility and equipment degradation at the residential consumers level, in: *2021 IEEE PES Innovative Smart Grid Technologies Europe (ISGT Europe),* 2021, pp. 1–6
10. M. Almassalkhi, J. Frolik, P. Hines, Packetized energy management: Asynchronous and anonymous coordination of thermostatically controlled loads, in: *2017 American Control Conference (ACC),* 2017, pp. 1431–1437